# *ARXON*: A Framework for Approximate Communication over Photonic Networks-on-Chip

Febin Sunny[1], Asif Mirza[1], Ishan Thakkar[2], Mahdi Nikdast[1], and Sudeep Pasricha[1]

[1]Colorado State University, Fort Collins, CO, USA [2]University of Kentucky, Lexington, KY, USA

[1]{febin.sunny, mirza.baig, mahdi.nikdast, sudeep}@colostate.edu [2]igthakkar@uky.edu

*Abstract*— The approximate computing paradigm advocates for relaxing accuracy goals in applications to improve energy-efficiency and performance. Recently, this paradigm has been explored to improve the energy-efficiency of silicon photonic networks-on-chip (PNoCs). Silicon photonic interconnects suffer from high power dissipation because of laser sources, which generate carrier wavelengths, and tuning power required for regulating photonic devices under different uncertainties. In this paper, we propose a framework called *ARXON* to reduce such power dissipation overhead by enabling intelligent and aggressive approximation during communication over silicon photonic links in PNoCs. Our framework reduces laser and tuning-power overhead while intelligently approximating communication, such that application output quality is not distorted beyond an acceptable limit. Simulation results show that our framework can achieve up to 56.4% lower laser power consumption and up to 23.8% better energy-efficiency than the best-known prior work on approximate communication with silicon photonic interconnects and for the same application output quality.

## I. INTRODUCTION

To match the increasing demand in processing capabilities of modern applications, the core count in emerging manycore systems has been steadily increasing. For example, Intel Xeon processors today have up to 56 cores [1], while NVIDIA's GPU's have reported over 8000 shader cores [2]. Emerging application-specific processors are pushing these numbers to new highs, e.g., the Cerebras AI accelerator has over 400,000 light weight cores [3]. The increasing number of cores creates greater core-to-core and core-to-memory communication.

Conventional metallic interconnects and electrical networks-on-chip (ENoCs) already dissipate very high power to support the high bandwidths and low-latency requirements of data-driven parallel applications today and are unlikely to scale to meet the demands of future applications [4]. Fortunately, chip-scale silicon photonics has emerged in recent years as a promising development to enhance ENoCs with light speed photonic links that can overcome the bottlenecks of slow and noise-prone electrical links. Silicon photonics can enable photonic NoCs (PNoCs) with a promise of much higher bandwidths and lower latencies than ENoCs [5].

Typical PNoC architectures employ several photonic devices such as photonic waveguides, couplers, splitters, and multi-wavelength laser sources, along with microring resonators (MRs) as modulators, detectors, and switches [5]. A laser source (either off-chip or on-chip) generates light with one or more wavelengths, which is coupled by an optical coupler to an on-chip photonic waveguide. This waveguide guides the input optical power of potentially multiple carrier wavelengths

(referred to as wavelength-division-multiplexed (WDM) transmission), via a series of optical power splitters, to the individual nodes (e.g., processing cores) on the chip. Each wavelength serves as a carrier for a data signal. Typically, multiple data signals are generated at a source node in the electrical domain as sequences of logical 0 and 1 voltage levels. These input electrical data signals are coupled with (i.e., modulated onto) the wavelengths using a group (bank) of modulator MRs (e.g., 64-bit data modulated on 64 wavelengths), typically using On-Off Keying (OOK) modulation. Subsequently, the carrier wavelengths are routed over the PNoC till they reach their destination node, where the wavelengths are filtered and dropped into the waveguide by a bank of filter MRs that maneuver the wavelengths to photodetectors to recover the data in the electrical domain. Each node in a PNoC can communicate to multiple other nodes through such WDM-enabled photonic waveguides in PNoCs.

Unfortunately, optical signals accumulate losses and crosstalk noise as they traverse PNoCs, necessitating high signal power from the laser for signal-to-noise ratio compensation and to guarantee that the signal can be received at the destination node with sufficient power to enable error-free recovery of the transmitted data. Moreover, the sensitivity of an MR to the wavelength it is intended to couple with is related to its physical properties (e.g., radius, width, thickness, refractive index of the device material) that can vary with fabrication and thermal variations. To rectify these problems, MRs must be "tuned" to correct the impact of variations either by free-carrier injection (electro-optic tuning) or thermally tuning the device (thermo-optic tuning). Such tuning entails energy and power overheads, which can become significant as the number of MRs in PNoCs increases. Novel solutions are therefore urgently needed to reduce these power overheads, so that PNoCs can serve as a viable replacement to ENoCs in emerging and future manycore architectures.

One promising direction towards this goal is approximate computing. As computational complexity and data volumes increase for emerging applications, ensuring fault-free computing for them is becoming increasingly difficult, for various reasons including: *(i)* increasing resource demands for big-data processing limit the resources available for traditional redundancy-based fault tolerance, and *(ii)* the ongoing scaling of semiconductor devices makes them increasingly sensitive to variations, e.g., due to imperfect fabrication processes. Approximate computing, which trades off "acceptable errors" during execution to reduce energy and runtime, is a potential solution to both these challenges [6]. With diminishing performance-per-watt gains from Dennard scaling, leveraging such aggressive techniques to achieve higher energy-efficiency is becoming increasingly important.

In this paper, we explore how to leverage data approximation to benefit the energy and power consumption footprints of PNoC architectures. To achieve this goal, we analyze how data approximation impacts the output quality of various applications, and how that will impact energy and power requirements for laser operation, transmission, and MR tuning. Our proposed framework, called *ARXON*, extends our previous work (*LORAX* [17]) to implement an aggressive loss-aware approximated-packet-transmission solution that reduces power overheads due to the laser, crosstalk mitigation, and MR tuning.

The novel contributions of this work are as follows:

1) We develop an approach that relies on approximating a subset of data transfers for applications, to reduce energy consumption in PNoCs while still maintaining acceptable output quality for applications;

2) We propose a strategy that adaptively switches between two modes of approximate data transmission, based on the photonic signal loss profile along the traversed path;

3) We evaluate the impact of utilizing multilevel signaling (pulse-amplitude modulation) instead of conventional on-off keying (OOK) signaling during approximate transfers for achieving even greater energy-efficiency;

4) We explore how adapting existing approaches towards MR tuning and crosstalk mitigation can help further reduce power overheads in PNoCs;

5) We evaluate *ARXON* on multiple applications and show its effectiveness over the best-known prior work on approximating data transfers over PNoC architectures;

## II. RELATED WORK

By carefully relaxing the requirement for computational correctness, it has been shown that many applications can execute with a much lower energy consumption and without significantly impacting application output quality. Some examples for approximation-tolerant applications that can save energy through this approach include audio transcoding, image processing, encoding/decoding during video streaming [7], [8], and big-data applications [9], [10]. The fast-growing repository of machine-learning (ML) applications represents a particularly promising target for approximation because of the inherent resilience to errors in most ML applications. As an example, it is possible to approximate the weights (e.g., from 32-bit floating-point to 8-bit fixed point) in convolutional and deep neural networks and with negligible changes in the output classification accuracy [24]. Many other approaches have been proposed for ML algorithm-level approximations [25]-[27]. With the introduction of ML applications into resource-constrained environments such as mobile and IoT platforms, there is growing interest in utilizing approximated versions of ML applications for faster and lower-energy inference [28].

In general, the approximate computing solutions proposed to date can be broadly categorized into four types based on their scope [11]: hardware, storage, software, and systems. The approximation of *hardware* components allows for a reduction in their complexity, and consequently their energy overheads [12]. For instance, an approximate full adder can utilize simpler approximated components such as XOR/XNOR based adders and pass transistor-based multiplexers [13], [14]. Additional reduction in circuit complexity and power dissipation can be enabled by avoiding XOR operations [15]. Techniques for *storage* approximation can include reducing refresh rates in DRAM [18], [19], which results in a deterioration of stored data, but at the advantage of increased energy-efficiency. Approaches for *software* approximation include algorithmic approximation that leverages domain specific knowledge [19], [20], [21]. They may also refer to approximating annotated data, variables, and high-level programming constructs (e.g., loop iterations), via annotations in the software code [22], [23]. At the *system* level, approximation involves modification of architectures to support imprecise operations. Attempts to design approximate NoC architectures fall under this category.

Several efforts have attempted to approximate data transfers over ENoC architectures by using strategies that reduce the number of bits or packets being transmitted to reduce NoC utilization, and thus reduce communication energy. An approximate ENoC for GPUs was presented in [29], where similar data packets were coalesced at the memory controller, to reduce the packets that traverse over the network. A hardware-data-approximation framework with an online data error control mechanism, which facilitates approximate matching of data patterns within a controllable value range, for ENoCs was presented in [30]. In [31], traffic data was approximated by dropping values from a packet before it is sent on to the ENoC, at a set interval. The data is then recreated at the destination nodes using a linear interpolator-based predictor. A dual voltage ENoC is proposed in [32], where lower-priority bits in a packet are transferred at a lower voltage level, which can save energy at the cost of possible bit flips. In contrast, the higher priority bits of the packet, including header bits, are transmitted with higher voltage, ensuring a lower bit-error rate (BER) for them. These approaches, e.g., [31], [32], focus on approximations for ENoCs. PNoCs utilize photonic links with very different mechanisms for data modulation, crosstalk, power dissipation, and signal losses, compared to ENoCs. Even though there are similarities in concepts used in photonic and electrical domains (e.g., lowering laser power swing can be considered analogous to lowering laser power) the design considerations, modeling, optimization strategies, and implementation in hardware required for a PNoC are very different from an ENoC. The design space of approximation techniques in PNoCs remains largely unexplored.

A recent work [16] explored the use of approximate data communication in PNoCs for the first time. The authors explored different levels of laser power for transmission of bits across a photonic waveguide, with a lower level of laser power used for bits that could be approximated, but at the cost of higher BER for these bits. The work focused specifically on approximation of floating-point data, where the least significant bits (LSBs) were transmitted at a lower laser-power level. However, the specific number of these bits to be approximated as well as the laser-power levels were decided in an application-independent manner, which ignores application-specific sensitivity to approximation. Moreover, the laser-power level is set statically and without considering the dynamic optical loss that photonic signals encounter as they traverse the network. In [17], we proposed *LORAX* framework that improved upon the work in [16] by using a loss-aware approach to adapt laser power at runtime for approximate communication in PNoCs. We analyzed the impact of adaptive approximation, varying laser-power levels, and the use of 4-pulse amplitude modulation (PAM4) on application output quality, to maximize application-specific energy savings in an acceptable manner.

The *ARXON* (<u>App</u>Ro<u>X</u>imation framework for <u>On</u>-chip photonic <u>N</u>etworks) framework presented in this article improves upon *LORAX* in multiple ways through: *(i)* considering integer data for approximation in addition to floating-point data (*LORAX* only considered floating-point data); *(ii)* integrating the impact of fabrication-process variations (PV) and thermal variations (TV) on MR tuning and leveraging it for energy savings; *(iii)* approximating error correction techniques, which are commonly used in PNoCs, to save more energy; and *(iv)* analyzing the potential for approximation for a much broader set of applications, and across multiple PNoC architectures. Section V describes our proposed *ARXON* framework in detail with evaluation results presented in Section VI.

## III. Data Formats and Approximations

### A. Floating-point Data

In many applications, floating-point data can be safely considered for approximation and without impacting the overall quality of the output from the approximation, as explored and demonstrated in [17]. The IEEE-754 standard defines a standardized floating-point data representation, which consists of three parts: sign (S), exponent (E), and mantissa (M), as shown in Fig. 1. The value of the data stored is:

$$X = (-1)^S \times 2^{E-bias} \times (1 + M) , \qquad (1)$$

Where X is the floating-point value. The *bias* values are 127 and 1203, respectively, for single and double precision representation, and are used to ensure that the exponent is always positive, thereby eliminating the need to store the exponent sign bit. The single precision (SP) and double precision (DP) representations vary in the number of bits allocated to the exponent and mantissa (see Fig. 1). E is 8 bits for SP and 11 bits for DP, while M is 23 bits for SP and 52 bits for DP. Also, S is 1 bit for both cases. From (1), we can observe that the S and E values notably affect the value of X. But X is typically less sensitive to alterations in M in many cases. M also takes up a significant portion of the floating-point data representation. We consider S and E as MSBs that should not be altered, whereas M makes up the LSBs that are more suitable for approximation to save energy during photonic transmission.

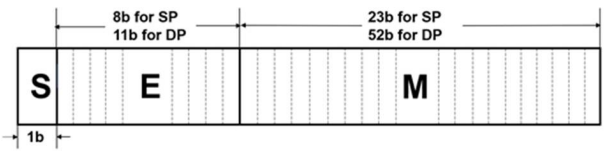

**Fig. 1: IEEE 754 floating-point representation.**

### B. Integer Data

It is more challenging to approximate integer data as it does not have a standard separation similar to the IEEE754 standard for floating-point data. An integer data value is usually represented as an N-bit chunk of data that can be signed or unsigned. If unsigned, the N-bits of data can be used to represent an integer value in the range from 0 to $2^N$-1. If signed, the most significant bit represents the sign bit, and the remaining N-1 bits represent an integer value in the range from

$- (2^{N-1}-1)$ to $+ (2^{N-1}-1)$. The number of bits, N, in an integer data word can change depending on the usage or application. N is usually in the range from 8 to 64 bits in today's platforms. Therefore, a generalized approach to approximate the integer data values is challenging. As a result, we have opted for an application-specific approach, where we identify possible integer variables that have larger than required size, depending on the values they handle. We deem the size of an unsigned integer variable as larger than required, if the most significant bits of the variable are not holding any useful information. We approximate such unnecessarily large unsigned-integer variables by truncating their MSBs. We also consider LSB approximation for integer packets, when viable. We found that integer data is generally not as tolerant to LSB approximation as floating-point data, so this approach cannot be as aggressive as LSB approximation in floating-point data and is thus used sparsely in our proposed framework.

### C. Applications Considered for Approximations

We evaluate the breakdown of integer and floating-point data usage across multiple applications, to establish how effective an approach that focuses on approximating floating-point LSB data and integer MSB data can be. We selected the ACCEPT benchmark suite [21], which consists of several applications, including some from the well-known benchmark suite PARSEC [33], that have been shown to have a relatively strong potential for approximations. While the applications in this suite may be executed on a single core, to adapt these to a PNoC-based multi-core platform with 64 cores, we used a multi-programmed simulation approach where the applications were replicated across the 64 cores to emulate parallel workloads on real multicore systems. Along with the applications from [21], we also considered several neural-network applications from the tinyDNN [34] benchmark suite to see how our approximation framework would fare for ML applications.

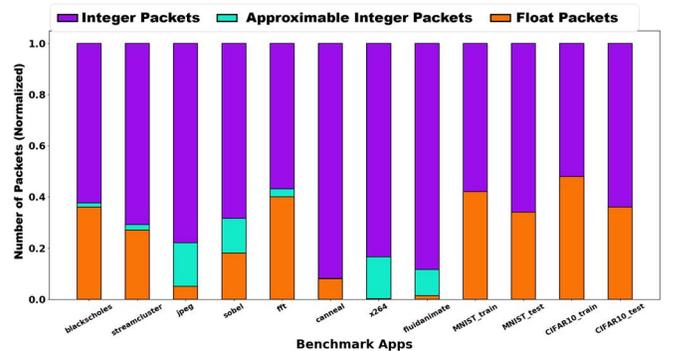

**Fig. 2: Characterization of applications considered for evaluation.**

We used the gem5 [35] full system-level simulator and the Intel PIN tool [36] in tandem to count the total number of integer and floating-point packets in transit across the memory hierarchy during the simulations. Fig. 2 shows the breakdown of the floating-point and integer packets across the applications for large input workloads. We considered all floating-point data packets as candidates for approximation. As for integer packets, we identified specific variables and a subset of their bits ("approximable integer packets") that can be approximated safely. The goal while selecting floating-point and integer packets for approximation was to keep application-specific error to below 10% of the original output. It can be observed that while a majority of the applications have integer packets that cannot be approximated without hurting output quality significantly, most of these applications have a non-trivial percentage of their packets that can be approximated. This is a promising observation that motivates our framework. But before we describe our proposed framework in detail (Section V), we briefly cover challenges in PNoCs related to crosstalk and signal loss (Section IV), which our approximation approach can leverage for energy savings.

## IV. CROSSTALK AND OPTICAL LOSS IN PNoCs

The overall data movement on the chip increases as the number of on-chip processing elements increases, and applications utilize more data. This requires a larger number of photonic waveguides, wavelengths, and MR devices to support the increased communication. However, using a larger number of photonic components makes it challenging to maintain acceptable BER and achieve sufficient signal-to-noise ratio (SNR) in any PNoC architecture due to signal optical loss and crosstalk noise accumulation in photonic building blocks [37].

Light propagation in photonic interconnects relies significantly on the precise geometry adjustment of photonic components. Any distortion in waveguide geometries and shape can notably impact the optical power and energy-efficiency in waveguides. For instance, sidewall roughness due to inevitable lithography and etching-process imperfections can result in scattering, and hence optical losses in waveguides [38]. In addition to such propagation loss, there is optical loss whenever a waveguide bends (i.e., bending loss), or when a wavelength passes (i.e., passing loss) or drops (i.e., drop loss) into an MR device. High optical losses require increased laser power to compensate for the loss and ensure appropriate optical-power levels at destination nodes where the signals are detected.

Crosstalk is another inherent phenomenon in photonic interconnects that degrades energy-efficiency and reliability. Crosstalk occurs due to variations in MR geometry or refractive index and imperfect spectral properties of MRs, which can cause an MR to couple optical power from another optical channel/wavelength (which acts as noise) in addition to its own optical channel (i.e., resonant wavelength). Such crosstalk noise is of concern in dense-wavelength-division multiplexed (DWDM) waveguides, necessary to support a higher bandwidth for emerging manycore platforms, where multiple optical channels exist with a small (e.g., <1 nm ) channel spacing. In such DWDM systems, not only will optical signals in each channel suffer from optical loss, but inter- and intra-channel crosstalk [40] accumulating on optical signals can severely reduce SNR and increase BER. Reducing crosstalk is challenging and techniques to minimize crosstalk (e.g., [41]-[43]) introduce further power and latency overheads.

It should be noted that the optical-power loss and crosstalk noise from a single silicon photonic device (e.g., MR) can be very small, and hence negligible [39]. However, in PNoCs integrating a large number of such devices (e.g., hundreds of thousands of MRs), the small power loss and crosstalk noise at the device-level accumulate to a point that they can severely reduce the performance and energy-efficiency in such architectures. In our proposed *ARXON* framework, as we are considering approximated data packets, we can intelligently relax crosstalk-mitigation mechanisms and optical loss compensation for the approximated bits, to aggressively reduce power and energy consumption overheads.

## V. *ARXON* FRAMEWORK: OVERVIEW

This section discusses the components of our *ARXON* framework. Section V.A provides an overview of our loss-aware laser power optimization strategy. Sections V.B and V.C discuss how crosstalk mitigation and tuning can be relaxed to save power during approximate-bit transfers. Lastly, Section V.D describes the integration of multilevel signaling to further reduce power dissipation during approximate communication in PNoCs.

### A. Loss-aware Laser Power Management for Approximation

Optical signals transmitted over a waveguide (photonic link) undergo attrition due to various optical losses they encounter along the path from a source to a destination, as discussed in section IV. To express how these optical losses tie in with the initial laser power provisioned to the optical signals in the waveguide, we can use the following model [51]:

$$P_{laser} - S_{detector} \geq P_{phot_loss} + 10 \times \log_{10} N_\lambda. \quad (2)$$

Here, $P_{laser}$ is the laser power in dBm, $S_{detector}$ is the receiver sensitivity, and $N_\lambda$ is the number of wavelength channels in the link. Also, $P_{phot\_loss}$ is the total optical loss accumulated on the optical signal during its transmission, which includes propagation, crossing, and bending losses in the waveguides, through- and drop-port losses of MR modulators and filters, and modulating loss in modulator MRs due to imperfect modulation [41]. $P_{laser}$ thus depends on the link bandwidth in terms of $N_\lambda$, and the total loss $P_{phot\_loss}$ encountered by each optical signal traversing the network. The $P_{phot\_loss}$ encountered along the network reduces the optical signal power. A signal can only be accurately recovered at the destination node if the received signal power is higher than $S_{detector}$. Ensuring this requires a high-enough $P_{laser}$ to compensate for all optical losses.

To approximate data transmission for floating-point data transfers, [16] used lower $P_{laser}$ for transmitting LSBs while keeping $P_{laser}$ unchanged for MSBs. However, if the destination node is relatively farther along a waveguide from a source node, the signals would encounter high losses and the signal power at the detector MRs would be lower than $S_{detector}$, which would result in detecting logic '0' for all the approximated signals at the destination node (e.g., with OOK modulation). In the scenario where the destination is closer to the source, it may be possible to detect the approximated signals accurately, as long as the losses encountered are low enough that the signal power at the detector MRs would be higher than $S_{detector}$, even with the reduced $P_{laser}$ for the approximated bits. For each data transfer on a waveguide, if we are aware of the distance of the destination from the source, it is possible to calculate the losses encountered for the signals, which can allow us to determine whether the signals can be recovered accurately, or if they will be detected as '0's. In such a scenario, it is more efficient to simply truncate all the approximated bits (i.e., reduce $P_{laser}$ to 0 for approximated signals) when the destination is farther along the waveguide and there is no likelihood of the signal being recovered accurately. Moreover, in the cases where the destination is closer to the source, we can transmit the approximated signals with a lower $P_{laser}$. This intelligent distance-aware transmission model for approximate data allows for some of the data to be detected accurately at the destination, while approximating other data depending on its content and distance to the destination.

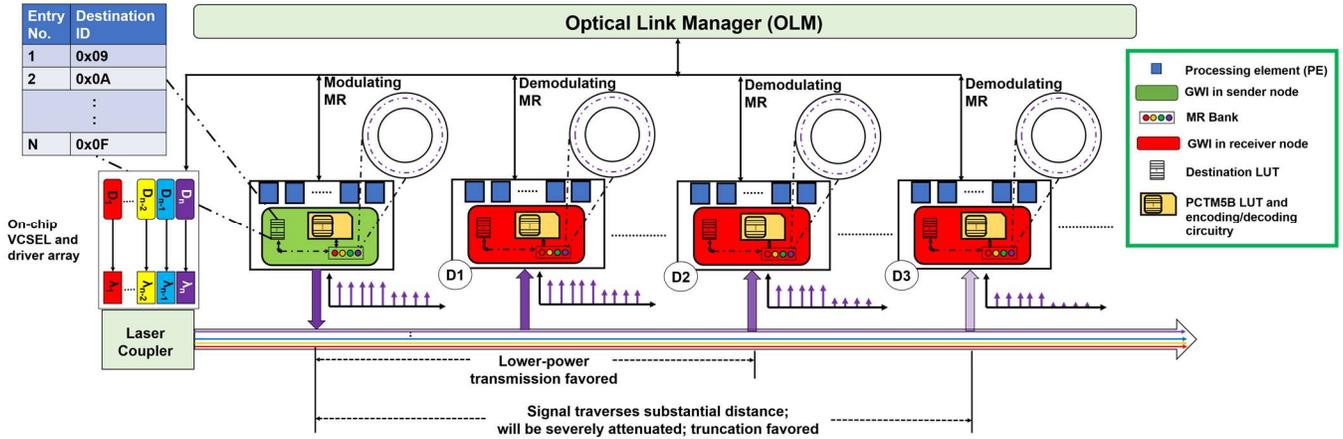

**Fig. 3: Overview of the proposed *ARXON* framework.**

Fig. 3 shows the operational details of the distance-aware transmission model in our framework, on a single-writer-multiple-reader (SWMR) waveguide that is part of a PNoC architecture. Note that while we illustrate our framework with an SWMR waveguide, our framework is also applicable (with minimal changes) to multiple-writer-multiple-reader (MWMR) and multiple-writer-single-reader (MWSR) waveguides that are also used in many PNoCs. In Fig. 3, only one sender node is active per data transmission phase and there is one receiver node (out of three in the figure) that is the destination for the transmission. In a pre-transmission phase (called receiver-selection phase), the sender notifies the receivers about the destination for the upcoming data transmission, and only the destination node will activate its MR banks, whereas the other nodes will power down their MR banks to save power in the transmission phase. As shown in Fig. 3, if the destination node is close to the sender node, (e.g., D1), we can transmit the approximated bit signals with a lower $P_{laser}$. Otherwise, if the destination node is farther away from the sender node, (e.g., D3), we determine that it would not be possible to detect the approximated signals at that destination due to the greater losses the signals will encounter. Therefore, we dynamically turn off $P_{laser}$, essentially truncating the bits.

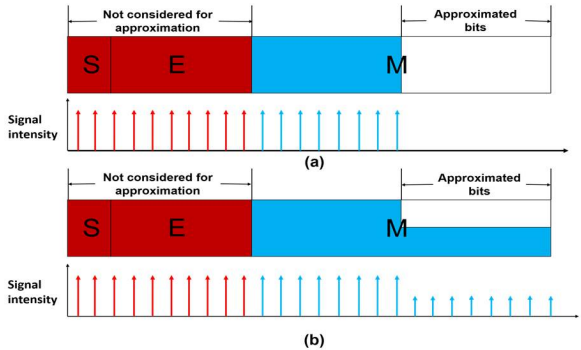

**Fig. 4: Floating-point data transmission on a photonic waveguide (a) truncation and (b) lower laser power.**

We consider both integer and floating-point data for approximation. For floating-point data, we perform distance aware transmission of the LSBs of the data in such a way that it will not impact the overall output quality of the application. Fig. 4 shows how transmission of data will conform to the distance aware transmission policy of our framework. In the case where substantial losses are expected to be encountered between a source and destination, we adopt the strategy shown in Fig. 4(a), where the data is truncated, as the approximated bits would have been lost during transmission anyway. When the data can have enough power to be successfully received at the destination node, we adopt the strategy shown in Fig. 4(b), where the data is transmitted at a lowered-laser power than its non-approximated counterparts. The power at which the bits can be transmitted, and the number of the approximated bits will depend on the application, as discussed in Section VI.

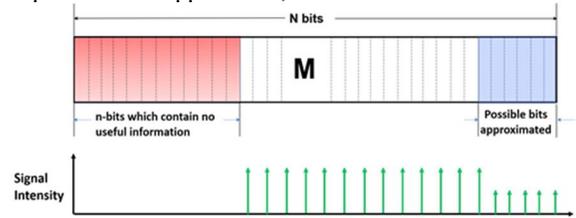

**Fig. 5: Approximated-integer-data transmission adopted.**

For approximating integer variables, we take a different approach. Based on our analysis, indiscriminate approximation to integer data in an application can significantly reduce output quality. Therefore, we instead profile applications and log the range of values stored in each integer variable. If the range of values is smaller than the bit size allotted to the variable (e.g., the case where a 32-bit integer variable only stores values up to 24 bits throughout the run of the application), we consider it a candidate for approximation. We can remove or truncate the MSBs that are unused in such variables that will otherwise take up modulation/demodulation and tuning energy necessary for transmission. We can also try and approximate the LSBs of the integer packets, and this approach can work in integer variables that store very large values where slight errors in the LSBs have minimal impact. But integer variables amenable to LSB approximation without significantly reducing output quality are rare. Nonetheless, for any such approximated LSB bits, the distance aware transmission model is applied as well. Fig. 5 summarizes our approximation strategy for integer packets.

To implement these strategies, we require: *(i)* a laser-control mechanism that can dynamically control the laser power being injected into the on-chip waveguides, and *(ii)* a mechanism to annotate approximable variables in the application source code, for runtime adaptation of transfers involving these variables.

We utilize an on-chip laser array with vertical-cavity surface-emitting lasers (VCSELs) [45], which can be directly controlled using on-chip laser drivers. With the laser drivers, we can control the power fed into each individual VCSEL, thus controlling the power of the laser output for a particular wavelength corresponding to that VCSEL. The gateway interface (GWI) that connects the electrical layer of the chip to the PNoC (see Fig. 3), communicates the desired $P_{laser}$ power level (including 0 for truncation) to the drivers, via an optical link manager, similar in structure to the one proposed in [46].

Identification of candidate packets to be approximated is done at the processing-element level, via source-code annotations [21], to generate necessary flags for data that is approximable. The main considerations while generating the flags for the packet, in our framework, are to allow for proper decoding of approximated or truncated packets at the destination. For this, two additional flags must be included in the packet header, at the processing-element level. The first (1 bit) flag indicates whether the approximable packet contains integer or float data and the second (1 bit) flag indicates whether the approximation is to be done for LSBs or MSBs. The number of bits that can be safely approximated or truncated are determined offline for each application and stored in lookup tables (LUTs) at the network interface (NI) which connects processing elements to routers that are in turn connected to GWIs. The number of bits approximated/truncated in a packet is also passed as part of the header flit of the packet to the GWI. This information can be used to gate (i.e., prevent) those bits from being passed into encoding/decoding circuitry. Note that as the number of bits truncated/approximated is necessary information for decoding, we must convey this information to the destination GWI as well. For this, we use six bits in the packet header. These six bits represent the number of approximated/truncated bits in the range from 0 to 32 bits, which is the range of approximation/truncation in our work.

Usually the header flit contains the routing information, which can just be the destination address. We consider a flit size of 64 bits, i.e., 64 bits are transmitted per transmission cycle. The number of used bits in the header flit do not exceed 16 bits (for the destination and source addresses), thus making it possible to incorporate the 8 necessary bits containing the two bits for the necessary flags and six bits for the approximation/truncation size information without causing any additional latency overheads. Once the header flit is received at the destination GWI, the flags and the approximated/truncated bits information are used to select the appropriate LSB/MSB to not be considered for decoding. The packet ID from the flit can then be used to track the remaining flits in the packet and treat them accordingly, if they were approximated/truncated.

Once the approximable bits have been identified, we must determine whether the approximation during their transfer is to be accomplished via reduced power transmission or truncation. This requires a LUT at each GWI (see Fig. 3) with the IDs of all the destination GWIs. The table at a source is populated with the destination IDs to which the loss values are sufficiently large enough to warrant truncation. The values can be easily calculated post-fabrication at design time, as the location of destination nodes as well as the cumulative loss to their GWI from the source does not significantly change at runtime. Once the decision to truncate or transmit at a lower laser power is made, depending on the destination node, the required power levels for the wavelengths are communicated to the VCSEL drivers via the optical-link manager. We discuss the overheads of the tables and the application specific $P_{laser}$ for the approximated signals in Section VI.

### B. Relaxed Crosstalk Mitigation Strategy

Due to the challenges with signal crosstalk outlined in Section IV, PNoCs must utilize one or more crosstalk mitigation strategies to reduce and achieve high SNR. We consider a state-of-the-art crosstalk mitigation strategy from [43] that can be applied at the link level in PNoCs. Analyses from [43] showed that a '1' carried by the wavelengths in the DWDM wavelength group adjacent to the resonant wavelength of an MR causes higher crosstalk in that MR. An encoding strategy was proposed to reduce inter-channel crosstalk noise by replacing instances of '1' values in adjacent wavelengths with '0' values, which helped reduce the optical signal-strength of immediate non-resonant wavelengths and improve SNR. Two encoding techniques were proposed that encoded nibbles (4-bits) of data. The PCTM5B technique encoded the nibble to 5-bit data, while the PCTM6B technique encoded the nibble to 6-bit data. Table 1 shows the code words used in these encoding techniques. Note that to implement PCTM5B on a photonic link with 64-bit word parallel transfers, 16 additional bits are required, which increases the number of MRs by 25%. Similarly, for PCTM6B, 32 additional bits are required for a 64-bit data word, and this increases the number of MRs by 50%. We assume that the lower-overhead PCTM5B technique is integrated into PNoCs by default, to meet BER goals.

**Table 1: Data word to code word conversion [43].**

| Code Words for PCTM5B Technique | | | |
|---|---|---|---|
| Data Word | Code Word | Data Word | Code Word |
| 0000 | 00000 | 1000 | 01000 |
| 0001 | 00001 | 1001 | 01001 |
| 0010 | 00010 | 1010 | 01010 |
| 0011 | 10101 | 1011 | 10100 |
| 0100 | 00100 | 1100 | 01100 |
| 0101 | 00101 | 1101 | 10010 |
| 0110 | 00110 | 1110 | 10001 |
| 0111 | 10110 | 1111 | 10000 |
| Code Words for PCTM6B Technique | | | |
| Data Word | Code Word | Data Word | Code Word |
| 0000 | 000000 | 1000 | 001000 |
| 0001 | 000001 | 1001 | 001001 |
| 0010 | 000010 | 1010 | 001010 |
| 0011 | 100000 | 1011 | 010100 |
| 0100 | 000100 | 1100 | 100010 |
| 0101 | 000101 | 1101 | 010010 |
| 0110 | 010101 | 1110 | 010001 |
| 0111 | 100001 | 1111 | 010000 |

In order to mitigate crosstalk, we assume the baseline configuration of the PNoC to implement PCTM5B. This means the encoder/decoder circuitry and the LUT, containing the data word-code word pairs, are incorporated into the GWI. Using these additions, the incoming packets from the processing elements can be encoded before they are transmitted to their destination, and at the destination, the packets are decoded using the LUTs. In our framework, applying crosstalk mitigation via PCTM5B technique to the truncated or approximated bits is an unnecessary overhead as it does not provide any benefits towards BER. By relaxing crosstalk mitigation for the truncated or approximated bits, it is possible to reduce the energy costs of the mitigation strategy. We do this by leveraging the approximation information gathered using our offline analysis of applications, where we consider that some LSB/MSB of the data can be approximated/truncated. During the encoding process, we do not consider these bits by gating their access to the encoder. Similarly, at the destination, when an approximated/truncated packet is received, the information from our LUTs are used to gate the approximated/truncated bits from being passed into the decoder circuitry.

### C. Relaxed MR Tuning Strategy

Thermal or electrical tuning of MRs in a PNoC is crucial for ensuring reliable communication, by counter-acting the effects of PV and TV. We assume the use of thermo-optic tuning in PNoCs, due to its better range of $\Delta\lambda_R$ correction. Electro-optic tuning can provide a tuning range of at most 1.5 nm [47]. In contrast, thermo-optic tuning can provide a tuning range of about 6.6 nm corresponding to the temperature range of up to 60K [44] at 0.11 nm/K sensitivity [48]. This comes at the price of higher energy consumption (~mW/nm) and slower operation (in units of μs). In our framework, we aim to reduce the overhead of tuning the MRs associated with truncated bits. We do not consider approximated bits for relaxed MR tuning, as the added noise this approach generates, due to thermal drift of $\lambda_R$, may render the approximated bits unreadable at the destination GWI. We do however relax the requirement for tuning MRs associated with the truncated bits, by temporarily turning off the tuning mechanism for those MRs.

### D. Integrating Multi-Level Signaling

The discussion in the previous sub-sections assumes the use of conventional on-off keying (OOK) signal modulation, where each photonic signal can have one of two power levels: high or on (when transmitting a '1'), and low or off (when transmitting a '0'). In contrast, multilevel signaling is a signal-modulation scheme where more than two levels of voltage can be used to modulate multiple bits of data simultaneously in each optical signal. The obvious benefit with such multilevel signaling is an increase in the bandwidth. Leveraging this technique in the photonic domain has, however, traditionally been a cumbersome process with high overheads, e.g., when using the signal superposition techniques from [49]. But with advances such as the introduction of Optical Digital to Analog Converter (ODAC) circuits [50] that are much more compact and faster than Mach-Zehnder Interferometers (MZIs) used in techniques involving superimposition [49], multilevel signaling has been shown to be more energy efficient than OOK [51], making it a promising candidate for more aggressive energy savings in silicon photonic networks.

Four-level pulse amplitude modulation (PAM4) is a multilevel signal modulation scheme where two extra levels of voltage (or optical signal power in case of optical modulation) are added in between the '0' and '1' levels of OOK. This allows PAM4 to transmit two bits per modulation as opposed to one bit per modulation in OOK. This in turn increases the bandwidth when compared to OOK. We are interested in evaluating the impact of using PAM4 in PNoCs and how its use will impact the effectiveness of our approximation strategies in *ARXON*. While PAM4 promises better energy-efficiency than OOK, it is prone to higher BER due to having multiple levels of the signal close to each other in the spectrum. Thus, we cannot reduce the laser power level of the LSB bits to the level used in OOK, as it would significantly reduce the likelihood of accurate data recovery even when destination nodes are relatively close to the source. Thus, when PAM4 is used, we need to increase the laser power compared to OOK. We used an empirically determined value of approximately 1.5× the laser power that was used for OOK, to prevent the degradation of approximated signals transmitted with PAM4. This may seem like a backward step in conserving energy, but the reduced-operational cost per modulation and the reduced-wavelength count for achieving the same bandwidth as OOK can reduce the overall laser power consumption. Also, while it is possible to add more signaling levels (e.g., to use a PAM8 modulation scheme [52]), as the number of amplitude levels increases, the optical signal becomes extremely susceptible to noise and causes increase in BER [53]. To ensure reliable communication when using PAM8, the bandwidth and speed of operation must be sacrificed [52]. Considering these constraints, we limit the extent of multilevel signaling integration in our framework to PAM4. The experimental results in the next section quantify the impact and trade-off when using PAM4 signaling in our framework.

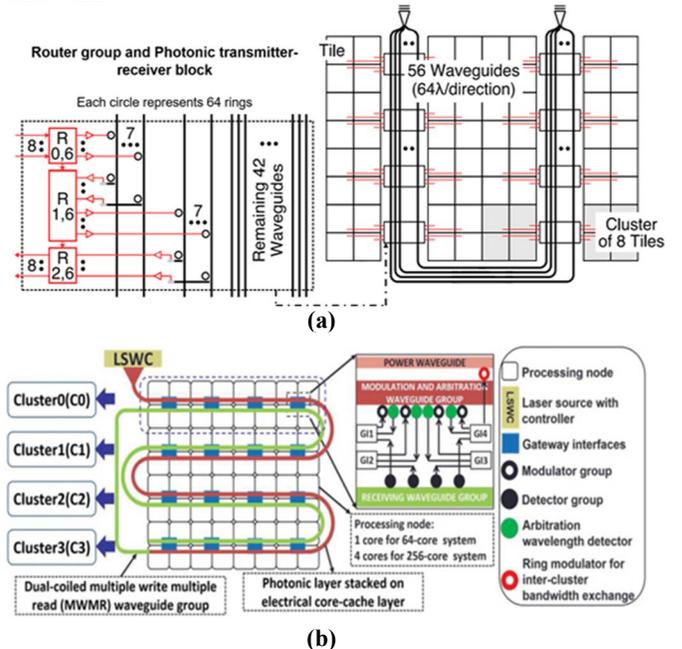

**Fig. 6: PNoC architectures considered for analyses. (a) Eight-ary three-stage Clos architecture with 64 cores [54]; and, (b) Schematic overview of SwiftNoC architecture [55].**

## VI. *ARXON* EVALUATION AND SIMULATION RESULTS

### A. Evaluation Setup

To evaluate our proposed *ARXON* framework, we implement it in Clos [54] and SwiftNoC PNoC architectures [55] for a 64 core processor, with baseline OOK signaling, PCTM5B crosstalk mitigation, and thermo-optic tuning in MRs.

The Clos PNoC, shown in Fig. 6(a), has an 8-ary 3-stage topology for a 64-core system with eight clusters and eight cores per cluster. It utilizes an optical crossbar topology with point-to-point photonic links utilizing SWMR waveguides for inter-cluster communication. Each cluster has two concentrators and a group of four cores is connected to each concentrator, where concentrators communicate with each other using an electrical router.

For the SwiftNoC PNoC, as shown in Fig. 6(b), we have again considered a 64-core system. Each node here has four cores and communication within the node happens through a 5×5 router, with the fifth port of the router connected to a GWI, which facilitates transfers between the CMOS-electrical layer and the photonic layer. Each GWI connects four nodes. The architecture utilizes eight waveguide groups with four MWMR waveguides per group in a crossbar topology. In order to support the MWMR communication, SwiftNoC utilizes a concurrent token stream arbitration that provides multiple simultaneous tokens and increases channel utilization.

The Clos PNoC has a waveguide length of 4.5$cm$ and the SwiftNoC PNoC has a waveguide length of 8.3$cm$ over the considered 400$mm^2$ chip. In both PNoCs, the first MR is encountered at ~1cm and the last MR is encountered at ~3.8$cm$ for Clos PNoC and ~7.8$cm$ for SwiftNoC. These distances have a key relationship to the laser power consumption, which we try to capture using the power model in (2). This relationship is visualized in Fig. 7, where the sudden jumps in power indicate a new GWI with the optical devices being encountered along the waveguide.

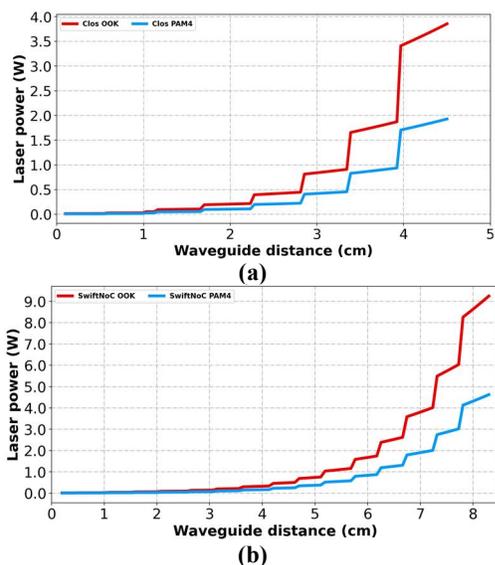

Fig. 7: Laser power consumption behavior over the length of the waveguide in (a) Clos PNoC and (b) SwiftNoC.

The considered PNoC architectures were modeled and simulated using an in-house SystemC based cycle-accurate simulator. A combination of gem5 full-system simulator [35] and Intel PIN toolkit [36] was utilized to generate traffic traces for the entire application. The traces were replayed on the PNoC simulators to determine energy savings in the PNoC. The PIN tool was used to obtain the addresses of the variables we deemed suitable for approximation from our analysis of applications and then to track accesses to them. Using this information in gem5 simulation, we track the relevant data flow at various levels of the simulated system (processor level, memory controller level, DRAM level, and cache level). The information generated while the simulation is running was consolidated and custom Python scripts were created to extract the necessary information about the data packets (e.g., timestamp at origin, their source, destination, data values, and control values from the packet header) and to generate the traces necessary for our cycle accurate simulator to simulate the applications on these PNoC architectures. Then, details of the approximate data communication (i.e., whether a packet was truncated or transmitted in lower power) was used to modify data in a subsequent gem5 simulation, to estimate the impact of the approximation on output quality for the application being considered. Table 2 shows gem5 architectural parameters considered in our experiments. We have based our simulations on x86 cores, but these simulations and our approach is applicable to systems having other types of cores as well, for e.g., ARM cores. Twelve applications from the ACCEPT and tinyDNN benchmark suites were used in our evaluations. The performance was evaluated at the 22 nm CMOS node for 400 $mm^2$ chips, with cores and routers operating at 5 GHz clock frequency. DSENT [56] was used to calculate the energy consumption of routers and the GWI at each node. Each GWI holds two LUTs for our framework; these are: one which holds the information regarding which destination addresses are preferred for truncation, and another for PCTMB5 encoding scheme. The size of both the LUTs at GWI level is fixed and is application independent. The PCTM5B LUT takes up only 144 bits for storing encoding decoding information at each GWI. The destination ID LUT can take up a maximum of 32 bits at each GWI for Clos PNoC and 64 bits for SwiftNoC variants.

**Table 2: 64-core architecture configuration.**

| Simulated component | Specification |
|---|---|
| No. of cores, processor type | 64, x86 |
| DRAM | 8GB, DDR3 |
| Memory controllers | 8 |
| L1 I/D cache, line size | 128KB each, direct mapped, 64B |
| L2 cache, line size, coherence | 2MB, 2-way set associative, 64B, MESI |

The table containing information regarding number of bits to be approximated/truncated for integer/float approximable packets is stored at the network interface (NI) of each processor. The maximum number of bits required in these LUTs for the worst case (application with the highest number of approximable variables) is a few hundred bits for the applications we considered. CACTI v6.5 [57] and scaling equations from [69] were used to evaluate the power, area, and delay for the lookup tables in NIs and GWIs. These values were found to be 0.236 $mm^2$ for the area consumption for all the tables, with a total power overhead, for reading from and writing into the tables, of 0.135 $mW$ for Clos and 0.472 $mm^2$ and 0.27 $mW$ respectively, for SwiftNoC. The combined power and area consumption of associated circuitry necessary for accessing information in the LUTs, calculated using gate-level analysis, is 0.0274 $mm^2$ and 4.224 $mW$ for Clos, and 0.0548 $mm^2$ and 8.448 $mW$ for SwiftNoC. LUTs in both Clos and SwiftNoC have the same number of entries as both architectures have the same number of processing elements. The encoding/decoding scheme is the same and the approximations done depend on the output error quality of the application and not the architecture, while SwiftNoC has double the number of GWIs. The access time for scratchpad RAMs designed with 22 nm technology node was under 1 cycle from synthesis estimates.

VCSEL control in *ARXON* was modeled after the optical link manager in [46], where the channel management for their PNoC design was described. However, since we are considering PNoCs from prior works with their own channel management systems in place for our analysis, we only adapt the approach for VCSEL control from [46]. The VCSEL control described in [46] uses a combination of MRs and PDs, but we only require the MR based switching mechanism for the VCSEL output. From the data available in [46] we calculated the area overhead necessary for implementing the VCSEL control, which was 0.093 $mm^2$ for OOK variants and 0.047 $mm^2$ for PAM4 variants of both the architectures.

**Table 3: Loss and power parameters**

| Parameters considered | Standard values | Aggressive values |
|---|---|---|
| Receiver sensitivity | -20 dBm [58] | -23.4 dBm [62] |
| MR through loss | 0.02 dB [59] | 0.02 dB [59] |
| MR drop loss | 0.7 dB [64] | 0.5 dB [60] |
| Propagation loss | 1 dB/cm | 0.25 dB/cm [66] |
| Bending loss | 0.01 dB/90° [63] | 0.005 dB/90° [61] |
| Thermo-optic tuning | 6.67 mW/nm [65] | 240 µW/nm [48] |

Clos and SwiftNoC PNoC architectures with PCTM5B are used as baselines for our analyses in this work. We have also considered a two-cycle overhead for PCTM5B encoding and decoding of the signals, as calculated in [43]. We considered $N_\lambda = 64$ for OOK, which would enable 64-bit transmission across a waveguide per cycle. For PAM4, we only need to consider $N_\lambda = 32$ to achieve the same bandwidth as with OOK modulation. Table 3 shows the energy values for losses and power dissipation in different photonic devices. We use a

"standard" set of values for these parameters from existing prototyping efforts, and a more "aggressive" set of values as per future projections from various research efforts. Our approach sacrifices reliability of approximated bits in floating point data and selected integer variable data, for EPB and laser power savings, as discussed in Sections V.B, V.C, and V.D.

We use the standard values for most of our simulations and use the aggressive values in Section VI.D. These values are used to calculate laser power from (2) and total power after considering tuning and lookup-table overheads. We consider a laser efficiency of 10% for our on-chip VCSELs, which is midway, the initial and worst-case efficiencies mentioned in [45]. We additionally consider a PAM4-induced-signaling loss of 5.8 dB in $P_{phot\_loss}$ for laser power calculations for PAM4 [51]. To compensate for the increased sensitivity of PAM4 to bit errors, we also consider laser-power levels that are 1.5× than those used for OOK signaling. For ensuring reliable communication, we have considered a BER of $10^{-9}$ in our designs. Lastly, we calculated application output error for the non-machine learning applications due to our approximation approach as:

$$Percentage\ (Output)\ Error = \frac{|approximated value - exact value|}{exact\ value} \times 100. \quad (3)$$

The "exact value" refers to the original output values, which can be a set of values presented in the output files, like in the case of *Blackscholes*, or it can be pixel values of output images/frames, like in the case of *JPEG*, *Sobel* or *X264*. The "approximated value" refers to the value of these outputs once the approximation approach is applied to the applications. For our analysis, we assume an error threshold of 10% output error, which was seen experimentally to be the limit at which the errors became apparent in the outputs of the majority of the applications [17]. For example, artifacts become noticeable in *JPEG* output as we cross the 10% error threshold. Thus, we want to ensure that none of the approximation strategies degrade output quality by more than 10%. For our machine-learning applications, we have considered the drop in accuracy to measure the impact of our framework and we have set the threshold as 10% drop in the accuracy.

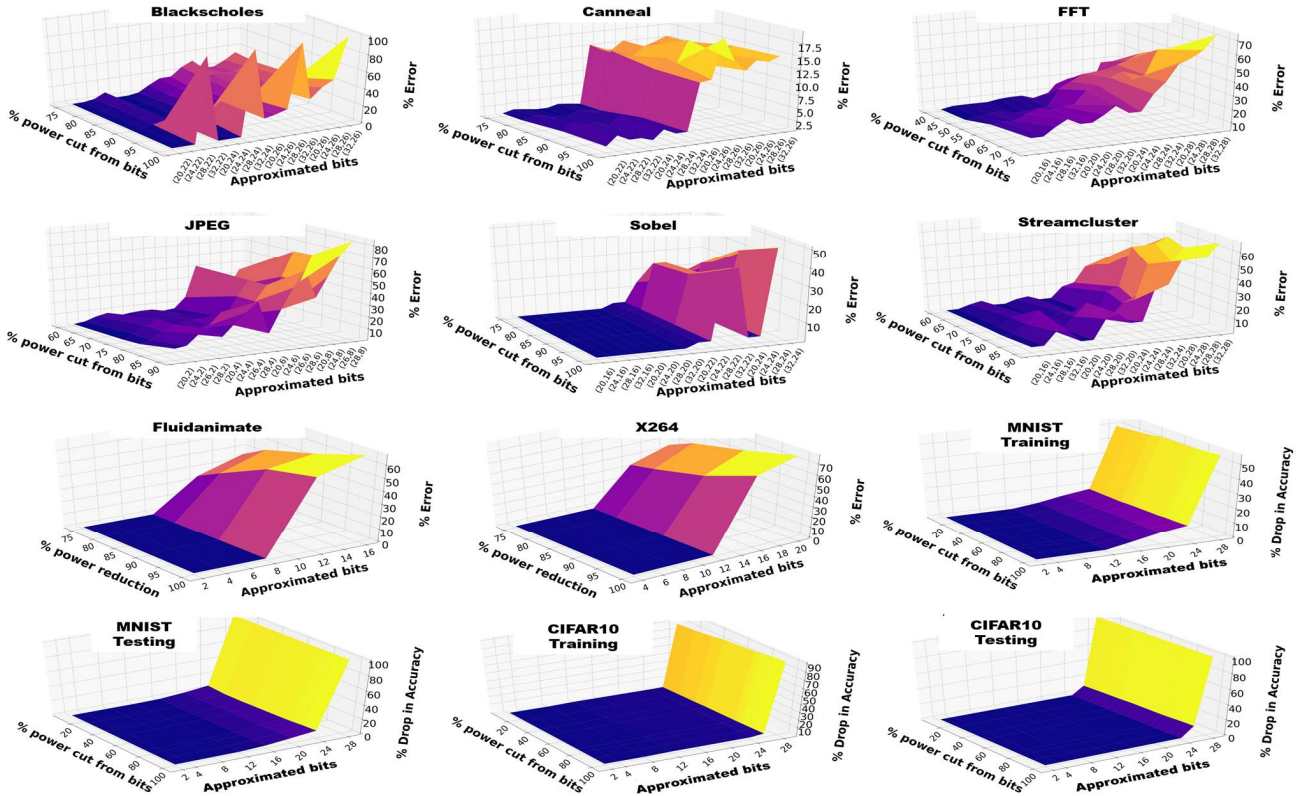

**Fig. 8:** Percentage error (PE)/Drop in accuracy in application output as a function of the number of approximated bit signals (y axis) and reduction in laser power (x axis) for the approximated signals, for blackscholes, canneal, fft, jpeg, sobel, streamcluster, fluidanimate, and X264 benchmarks with large input workloads and MNIST (training and testing) and CIFAR10 (training and testing) models.

### B. Impact of ARXON on Applications Considered

Our first set of experiments involve analyzing the sensitivity of an application to varying degrees of approximation of their floating-point data. We are interested in studying the impact on output error from approximating a number of bits in the packets carrying data deemed approximable. Additionally, we are also interested in studying the impact on output error of varying levels of lowered laser power for those approximated bits.

Fig. 8 shows the results of our comprehensive study for the applications we considered (as depicted earlier in Fig. 2). The z-axis shows the percentage error (PE) in application output, or drop in accuracy for ML applications, as a function of the reduction in $P_{laser}$ level for the photonic signals that carry the approximated bits (x-axis; varying from 0% to 100%, where 100% refers to truncation), and the number of bits that were considered for approximation (y-axis; with the number of approximated float and integer bits given in [float, integer] format). The subset of combination of these values were selected for enabling viable trade-offs between output quality and power consumption. It should be noted that not all applications consider both floating-point and integer data for approximation. For example, *Fluidanimate* only considers integers for approximation while the ML applications (*CIFAR10* and *MNIST*) only considers floating-point data. This selection of datatypes to be approximated was made after profiling the application and determining the datatypes that do not have adequate impact on the traffic (e.g., floating-point data in the case of *Fluidanimate* and *X264*) or the functionality of the application (integers in the case of the ML applications considered). This is a more comprehensive version of the experiments in our earlier work, presented in [17]. In those experiments in [17] we had determined how much floating-point approximation can be tolerated by the applications from ACCEPT benchmark. Here we not only consider a larger number and variety of applications, but also use more comprehensively determined thresholds than in [17] to explore how approximating the integer bits along with the float bits affects the output quality. It is clear from our analyses that not all applications can tolerate the same level of approximation. From the PE values, we can observe that *FFT* with a large volume of floating-point data traffic (see Fig. 2) reaches the error threshold of 10% rather quickly as the number of approximated bits increases and laser power-levels reduce, whereas *Canneal* with a lower floating-point traffic-volume observed seems to have very low PE values across the various experiments. The edge detection algorithm *Sobel* performs well in approximated conditions, possibly owing to the lowered data accuracy requirements to construct the output. *Streamcluster* involves an approximation strategy for data streams and is also observed to be quite resilient to greater levels of approximation. *Blackscholes*, which performs market options calculations is particularly sensitive to the approximated number of bits and the laser-power levels. *JPEG* performs image compression, and the output image quality is also more sensitive to approximation. *Fluidanimate* generates a video of flowing liquid depending on the input data provided. *X264* is a video codec, which generates compressed video from the input, which is raw video data. *Fluidanimate* and *X264* applications were subjected to only integer MSB approximation, and threshold is quickly breached after the amount of MSBs approximated start taking up bits which contain values, the quick rise in error can be explained by the fact that we are approximating MSBs which would cause very large shift in values. Moreover, we considered implementations of deep convolutional neural networks for classification of *CIFAR-10* and *MNIST* datasets, from tinyDNN. The machine learning applications used single precision floats, and we were able to approximate till the point where we encroached on the exponent, but the decay of output accuracy ramped up very quickly once we tried to approximate any further. From Fig. 8 we can see that there is a sharp increase in percentage output error (PE), as we approximated beyond a certain number of bits, in the case of many of the applications considered, e.g., the applications in the bottom two rows. The erratic jumps in error rate for the six applications in the top two rows of Fig. 8 are because we are considering discrete combinations of approximated bits for floating point and integer variables, along the 'Approximated bits' axis.

**Table 4: Number of bits considered for approximation and laser-transmission-power level for the corresponding signals across benchmarks and frameworks considered.**

| Application Name | Truncation | [16] | *LORAX* [17] | | *ARXON* (proposed) | | |
|---|---|---|---|---|---|---|---|
| | Truncated Bits (float) | | Approximated Bits (float) | % Power reduction | Approximated bits in floating-point packets | Approximated bits in integer packets | % Power reduction |
| **Blackscholes** | 12 | | 32 | 90 | 32 | 24 | 90 |
| **Canneal** | 32 | | 32 | 100 | 32 | 24 | 100 |
| **FFT** | 8 | | 32 | 50 | 32 | 20 | 50 |
| **JPEG** | 20 | | 24 | 80 | 22 | 4 | 80 |
| **Sobel** | 32 | 16 bits approximated, with 20% power reduction | 32 | 100 | 32 | 20 | 100 |
| **Streamcluster** | 12 | | 28 | 80 | 28 | 20 | 80 |
| **Fluidanimate** | - | | - | - | - | 8 | 100 |
| **X264** | - | | - | - | - | 12 | 100 |
| **MNIST_train** | 24 | | 24 | 100 | 24 | - | 100 |
| **MNIST_test** | 24 | | 24 | 100 | 24 | - | 100 |
| **CIFAR10_train** | 24 | | 24 | 100 | 24 | - | 100 |
| **CIFAR10_test** | 24 | | 24 | 100 | 24 | - | 100 |

Table 4 summarizes the best combination of approximable bits and the laser-power-transmission levels for these bits and for each application while ensuring that the application output error does not exceed 10% for our proposed framework (*ARXON*). Table 4 also shows the number of bits that can be truncated, selected to meet the <10% PE constraint. For the approach in [16], we perform approximations on 16 LSBs transmitted at 20% laser power (advocated as an optimal choice in that work), which also satisfies the <10% PE constraint.

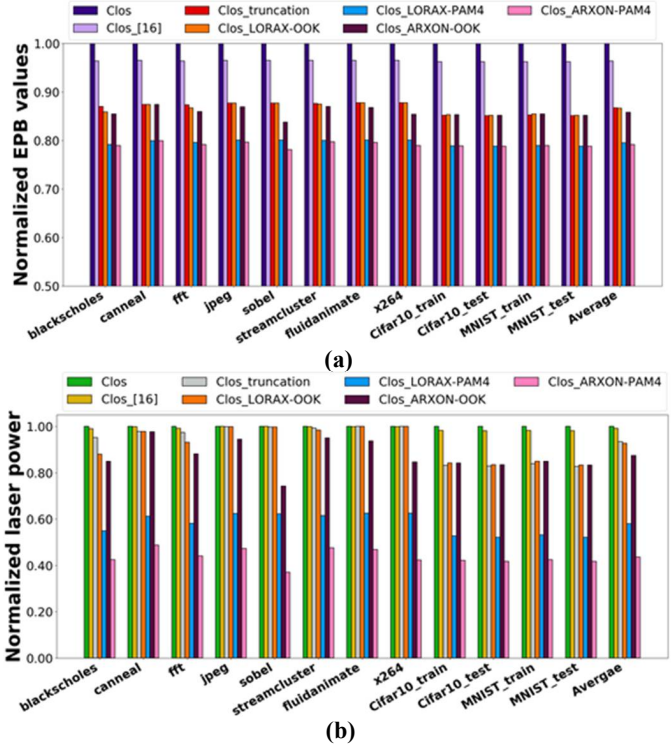

**Fig. 9: (a) Energy-per-bit (EPB) and (b) laser power comparison across different frameworks for Clos PNoC architecture.**

Fig. 9 shows the EPB and laser power comparison results for the various frameworks in the Clos PNoC architecture These analyses consider the benefits from distance-aware transmission and the relaxed encoding technique for approximated packets for *ARXON*. Fig. 9(a) shows that using *ARXON-OOK* results in lower EPB than the previous approaches, including our previous framework *LORAX-OOK*. The better EPB for *LORAX* and *ARXON* can be attributed the fact that they avoid wasteful transmission at lower laser power when it is unlikely that the destination can recover the transmitted data due to high optical losses. Also, [16] has noticeably higher EPB values for which we are not considering the benefits of relaxed encoding and distance-aware transmission for the framework to be consistent with the framework presented in that paper. The *ARXON-OOK* framework improves upon *LORAX*-OOK, [16] and truncation, by adaptively switching between truncation and an application-specific laser-power-intensity level for approximated bits of both floating-point and integer packets. The *ARXON-PAM4* variant of our framework achieves the largest reduction in EPB, even though it uses 1.5× higher laser-power levels for the approximated bits. The use of fewer wavelengths in PAM4 allows for more energy savings, despite greater losses and the use of more laser power per wavelength than OOK variant.

On average, *ARXON-PAM4* shows 21%, 17.2%, 9.7%, 9.2%, and 1.2% lower EPB compared to the baseline Clos, [16], truncation, *LORAX-OOK,* and *LORAX-PAM4* approaches, respectively. *ARXON-OOK* exhibits lower EPB on average while having a 6% higher EPB than the *LORAX-PAM4* approach. In the best case scenarios for the *Blackscholes* and *Sobel* applications, *ARXON-PAM4* has 21.2% and 23.5% lower EPB than the Clos baseline; and 17.4% and 15.6% lower EPB than [16]; 9.8% and 11.5% lower EPB when compared to truncation; 8.6% and 10.25% lower EPB than *LORAX-OOK,* and 1.24% and 2.5% lower EPB than *LORAX-PAM4* for these two applications.

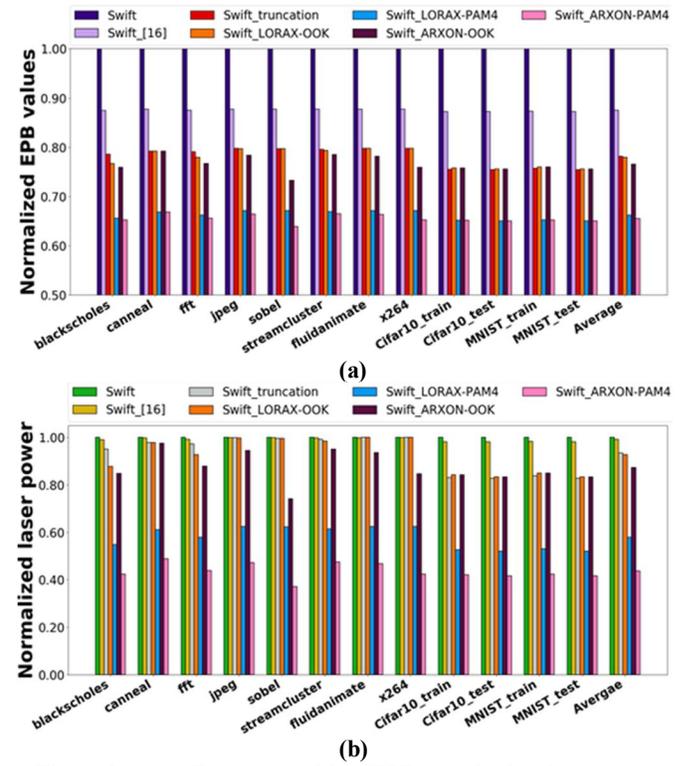

**Fig. 10: (a) Energy-per-bit (EPB) and (b) laser power comparison across different frameworks for SwiftNoC architecture**

Fig. 9(b) shows the laser power reduction. On average, *ARXON-PAM4* uses 50.45%, 49.5%, 43.2%, 42.5%, and 7.7% lower laser power compared to the baseline Clos, [16], truncation, *LORAX-OOK,* and *LORAX-PAM4,* respectively. *ARXON-OOK* exhibits lower average laser-power consumption on average while exhibiting 28% higher laser power consumption than *LORAX-PAM4*. For the best case *Blackscholes* and *Sobel* applications, laser power for *ARXON-PAM4* is 51.7% and 59.2% lower than the Clos baseline and 50.8% and 57.9% lower than [16], while against truncation it is 51% and 58.5% lower, against *LORAX-OOK* we see 38% and 57% lowered laser-power utilization and against *LORAX-PAM4* we have 6.5% and 20% lower laser-power utilization.

Fig. 10 shows the same analyses but done for the frameworks implemented on the SwiftNoC architecture. The larger data rate and the larger number of GWIs in the architecture has impacted the packets and their distance aware transmission profile, creating more avenues to truncate the packets, yielding better EPB results in this architecture. The general trend in EPB and laser-power savings is similar to that for the Clos architecture, with *Blackscholes* and *Sobel* applications again exhibiting the best EPB and laser-power saving values. From Fig. 10(a), *ARXON-PAM4* exhibits 36%, 23.8%, 13.5%, 12.9%, and 1.8% lower EPB on average than baseline SwiftNoC, [16], truncation, *LORAX-OOK*, and *LORAX-PAM4*, respectively.

The results for SwiftNoC show the same trend as the Clos architecture for normalized laser power (Fig. 10(b)), albeit with lower laser power across applications with average laser power consumption for *ARXON-PAM4* at 57.2%, 56.4%, 50.8%, 49.3%, and 15.7% better than baseline SwiftNoC, [16], truncation, *LORAX-OOK*, and *LORAX-PAM4*, respectively.

These results highlight the promise of our *ARXON* framework, as it improves upon the ability *LORAX* exhibited to trade-off output correctness with energy-efficiency and laser-power savings in PNoC architectures executing selected applications.

### C. MR Tuning Relaxation-Based Analyses

In addition to distance-aware transmission for float and integer packets and relaxing crosstalk-mitigation encoding techniques, we also consider the potential for relaxed thermo-optic tuning for truncated bits. We have considered thermal MR tuning in our work for its larger range of operation over other tuning methods such as electro-optic tuning. However, thermal tuning strategies are much slower in operation when compared to electro-optic tuning (microseconds for operation as opposed to nano to picoseconds for electro-optic tuning). But, this overhead cannot be avoided, as using just the electro-optic tuning method will not offer sufficient coverage for the thermal and process variations encountered by MRs, the effect of which must be mitigated for correct operation.

But with the increasing maturity of silicon photonics, we envision faster thermo-optic tuning strategies or a combination of different tuning strategies to reduce this tuning latency. Therefore, in this section we explore the potential of energy savings due to relaxed MR tuning, i.e., by turning off the tuning mechanism for MRs associated with truncated bits. For this experiment, we utilize thermal and process-variation information. For thermal variations (TV), we have referred to the study conducted in [67] and have adopted the worst-case TV induced shift to be 6.5 nm. For analysis of process variations (PV), we utilized the PV analysis method as described in [67], where PV is considered as a Gaussian random distribution. As the granularity of the method is at 30 nm, we have opted for analyzing PV at the GWI level rather than for individual MR devices. We have generated PV maps for the architectures using the method from [68] and have selected locations corresponding to the GWIs in the layouts. We took the average of device variations (i.e., width and thickness) in that location. This was repeated over 100 different PV maps.

Utilizing the PV and TV information obtained, we implement the tuning-relaxation approach, where we turn off thermo-optic tuning for all truncated bits. In order to implement the control necessary for relaxing the tuning, which in our case is to turn off tuning mechanisms to the MRs of truncated bits, we use a gating mechanism similar to the one utilized for the encoding strategy, as mentioned in Section V.B. With this mechanism, we can power gate the tuning circuits to the MR, as per the information from LUTs, again similar to the description in Section V.B. From our analysis, this had a substantial impact on the EPB values of our *ARXON* framework, as shown in Fig. 11. Our observations in Fig. 11(a) for Clos PNoC and Fig. 11(b) for SwiftNoC, show that the *ARXON* variants have substantial savings over the other frameworks considered, a trend maintained even while using the aggressive values as it was with standard values. This is because the tuning based approach is again dependent on the traffic profile of the applications, with higher truncated packets meaning better savings. So, we see *Blackscholes* and *Sobel* as the best performing applications again. We do not consider laser-power savings in this scenario, as the tuning relaxation approach does not impact the laser power. On average, *ARXON-PAM4* has 38.1%, 36.1%, 26.8%, 26.4%, and 19.2% better EPB values than baseline Clos, [16], truncation, *LORAX-OOK* and *LORAX-PAM4*. When implemented in SwiftNoC, *ARXON-PAM4* exhibits 48.6%, 39.3%, 29%, 28.5%, and 16.9% better EPB than baseline, [16], truncation, *LORAX-OOK* and *LORAX-PAM4*, respectively. This only adds to the significant reduction in the overall laser power consumption achieved by *ARXON*, showing how our framework achieves better laser power and EPB values for all the applications considered in our analyses.

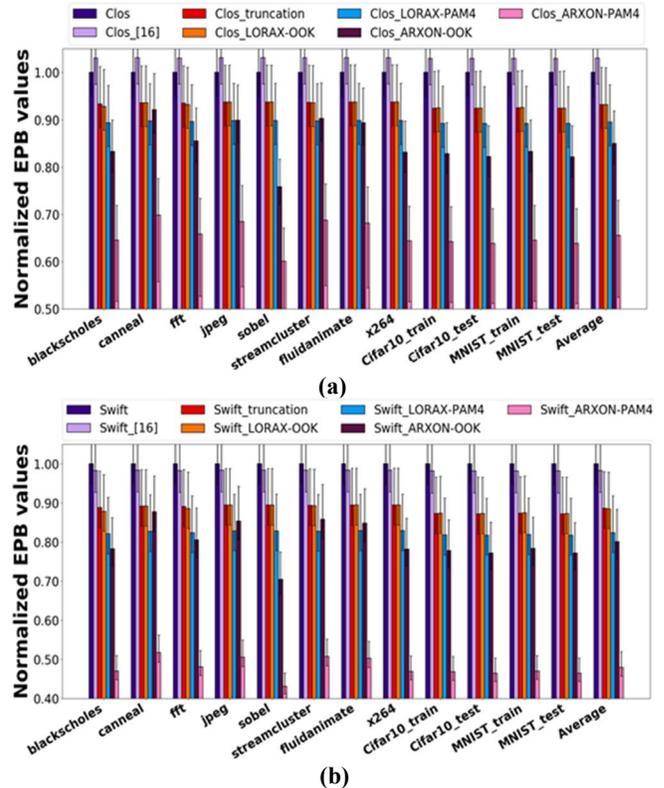

**Fig. 11:** EPB values for *ARXON* implemented on (a) Clos and (b) SwiftNoC while considering thermal-tuning relaxation.

### D. Power Dissipation Breakdown

We performed an experiment to determine how much more power can be saved as silicon photonics technology matures and devices with improved characteristics become available. For this, we contrast the power dissipation with our framework on the Clos and SwiftNoC architectures, for the standard and aggressive values of parameters in Table 3. As the EPB and laser power once normalized follows the same trends, we decided to use a detailed power-dissipation breakdown to show how much *ARXON* improves the power consumption in PNoC and in which areas.

Fig. 12 shows the detailed power breakdown for the frameworks, averaged across the applications. From the figures we can clearly observe how *ARXON* impacts both laser power and tuning-power dissipation, having the lowest power dissipation in both those categories and in total, be it while considering standard loss and power utilization values or while considering aggressive values.

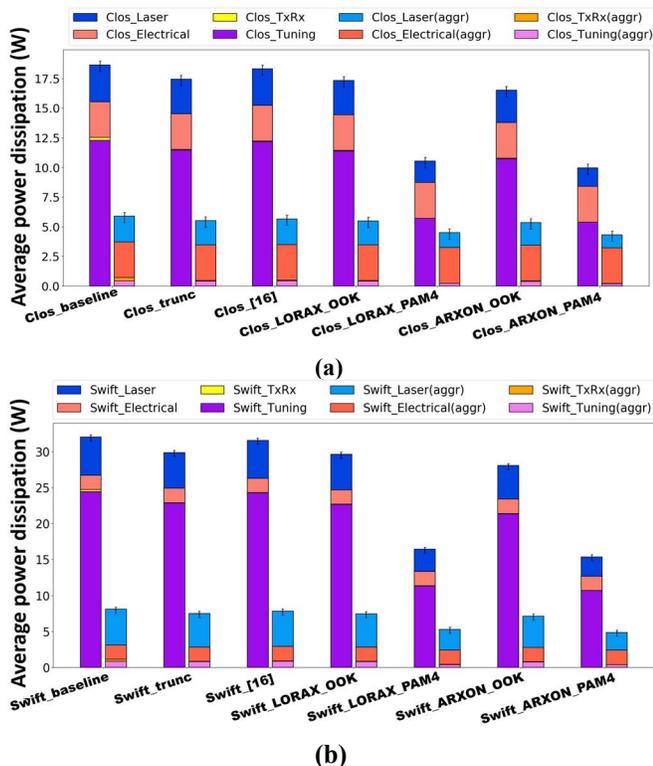

**Fig. 12: Power dissipation breakdown for standard and aggressive values ('aggr' in the plots) for (a) Clos and (b) SwiftNoC PNoCs.**

## VII. CONCLUSION

In this paper, we proposed a new framework called *ARXON* for loss-aware approximation of data communicated over PNoC architectures. We also studied how multilevel signaling can assist with the proposed approximation framework. We considered MR tuning and crosstalk mitigation strategies as avenues to save energy while our distance aware transmission technique is in effect. Our results indicate that using multilevel signaling as part of our framework can reduce laser-power consumption by up to 57.2% over a baseline PNoC architecture. Our framework also shows up to 56.4% lower laser power and up to 23.8% better energy-efficiency compared to the best-known prior work on approximating communication in PNoCs. These results highlight the potential of approximation in PNoC architectures to reduce energy and power consumption in emerging manycore platforms.

## ACKNOWLEDGEMENTS

This research was supported by the National Science Foundation (NSF) under grants CCF-1813370, CCF-2006788.

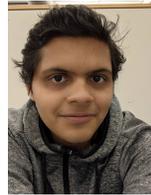

**Febin P Sunny** (S'19) is currently a PhD candidate at Colorado State University. His research interests include embedded systems, photonic networks on-chips and neuromorphic computing.

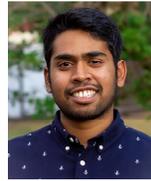

**Asif Mirza** (S'19) is currently a PhD candidate at Colorado State University, with research interests spanning device level silicon photonics and high performance computing.

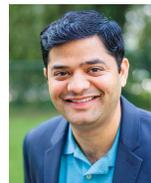

**Ishan Thakkar** (M'18) is an Assistant Professor in the Depart. of ECE of the University of Kentucky, Lexington, KY, USA. He received the MS and PhD degrees in EE from Colorado State University, USA in 2013 and 2018 respectively. His research interests include DRAM systems, non-volatile memories, in-memory computing, and optical networks-on-chip.

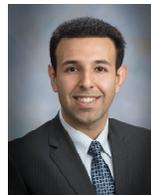

**Mahdi Nikdast** (M'14, SM'19) received his Ph.D. in ECE from the Hong Kong University of Science and Technology in 2013. He is currently an Assistant Professor of ECE at Colorado State University. His research interests include silicon photonics, high-performance computing, and emerging hardware technologies. He is a Senior Member of IEEE.

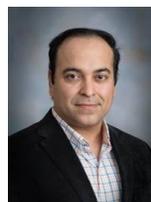

**Sudeep Pasricha** (M'02, SM'13) received his Ph.D. from the University of California, Irvine, USA, in 2008. He is currently a Professor of ECE at Colorado State University. His current research interests include energy-efficient and fault tolerance manycore computing, with an emphasis on silicon photonics, networks-on-chip, and machine learning. He is a Senior Member of IEEE.